# A read-filtering algorithm for high-throughput marker-gene studies that greatly improves OTU accuracy

Fernando Puente-Sánchez[1*], Jacobo Aguirre[1,2,3], Víctor Parro[1].

[1]Department of Molecular Evolution, Centro de Astrobiología (INTA-CSIC). Instituto Nacional de Técnica Aeroespacial, Ctra de Torrejón a Ajalvir km 4. 28850 Torrejón de Ardoz, Madrid, Spain.

[2]Centro Nacional de Biotecnología (CSIC). c/ Darwin 3, 28049 Madrid, Spain.

[3]Grupo Interdisciplinar de Sistemas Complejos (GISC).

[*]Correspondence should be addressed to F.P.S. (puentesf@cab.inta-csic.es).

## Abstract

Adequate read filtering is critical when processing high-throughput data in marker-gene-based studies. Sequencing errors can cause the mis-clustering of otherwise similar reads, artificially increasing the number of retrieved Operational Taxonomic Units (OTUs) and therefore leading to the overestimation of microbial diversity. Sequencing errors will also result in OTUs that are not accurate reconstructions of the original biological sequences. Herein we present a novel and sensitive sequence filtering algorithm that minimizes both problems by calculating the exact error-probability distribution of a sequence from its quality scores. In order to validate our method, we quality-filtered thirty-seven publicly available datasets obtained by sequencing mock and environmental microbial communities with the Roche 454, Illumina MiSeq and IonTorrent PGM platforms, and compared our results to those obtained with previous approaches such as the ones included in mothur, QIIME and UPARSE. Our algorithm retained substantially more reads than its predecessors, while resulting in fewer and more accurate OTUs. This improved sensitiveness produced more faithful representations, both quantitatively and qualitatively, of the true microbial diversity present in the studied samples. Furthermore, the method introduced in this work is computationally inexpensive and can be readily applied in conjunction with any existent analysis pipeline.

## Introduction

High-throughput sequencing of marker genes, such as the 16S ribosomal RNA, has become an invaluable tool for microbial ecologists, since it allows for a previously unreachable level of detail in the analysis of complex microbial communities. Many studies have used platforms such as the Roche 454, Illumina or IonTorrent sequencers to thoroughly characterize and compare microbial communities at an affordable cost (1, 2, 3, 4, 5), while others have taken advantage of their very high yield in order to analyse the structure and composition of the rare biosphere (6). However, the correct assessment of sequencing artefacts is critical in obtaining representative results. In whole genome sequencing studies an erroneous base can be corrected by overlapping reads during consensus sequence assemblage, but in marker-gene studies each read is assumed to come from a different individual in the community. In this case, sequencing errors can cause the mis-clustering of otherwise similar reads, resulting in the overestimation of microbial diversity (7). The most common software tools and packages include sequence clustering into OTUs in their recommended pipelines (8, 9, 10, 11, 12, 13, 14, 15, see 15 for a comparison of several molecular ecology pipelines). Alternatives to traditional clustering have been recently proposed, such as distribution-based clustering (16) or a clustering-free approach (17). These novel methods are specially suited for subpopulation level studies, but work only for moderate-to-high abundance sequences, being unsuitable for population-level alpha or beta diversity studies. Moreover, even although they can remove likely-erroneous sequences and resolve subpopulations based on dynamic information, they nevertheless rely on a quality-filtering step for the pre-processing of raw reads.

Amplicon denoising (18, 19) is a widespread method for filtering Roche 454 pyrosequencing reads that can also be applied to IonTorrent data. It works on flowgrams rather than sequences, which allows for a more natural modelling of the homopolymer read errors that are characteristic of pyrosequencing and ion semiconductor sequencing. However, it is platform specific and computationally expensive.

For Illumina systems there is no consensus approach to quality-filtering, with the authors of mothur (20), QIIME (21) and UPARSE (15) proposing different solutions. All those heuristic approaches were published as parts of their respective pipelines, but to the best of our knowledge they have not been thoroughly compared to each other.

The lack of a rigorous method for incorporating quality scores in the analysis of marker-gene sequences has also led some authors to advocate for a stringent filtering in order to reduce the retrieval of spurious diversity (3). However, over-stringent filtration will result in an undesired loss of sensitivity and will have an impact on the observed taxonomic distribution (21). Therefore, an accurate algorithm that overcomes these problems is desirable.

Herein we present and validate the *Poisson binomial filtering* (PBF) method, which is able to determine the exact error-probability distribution of any sequence with associated quality scores, by using a simple statistical approach. Phred quality scores, which represent the probability that a given base call is mistaken, can be derived from the raw output of every sequencing platform. Reading a single base can be likened to tossing a coin: the base is either right or wrong, and both chances can be determined from its quality score. In fact, the number of errors present in a given base follows a Bernoulli distribution, i.e. a binomial distribution with a single trial. For a sequence of nucleotides with different error probabilities, we sum their associated Bernoulli random variables in order to obtain the exact probability that the sequence has accumulated more than $k$ errors, where $k$ is the maximum number of errors that still allows for a correct clustering (**Supplementary Note 1)**. When compared with the filtering approaches included in mainstream molecular ecology pipelines such as mothur, QIIME or UPARSE, Poisson binomial filtering proved to be the most accurate algorithm for filtering marker-gene sequences. Additionally, PBF is based on simple statistical principles and, since it only requires Phred quality scores as an input, it is expected to work robustly regardless of the sequencing platform. Finally, our algorithm is computationally efficient, scales linearly with the number of sequences, and has a low memory fingerprint, making it useful even in low-performance desktop environments.

## Materials and methods

*The Poisson binomial filtering algorithm*

Let us suppose we have 1 sequence of length $N$ nucleotides (nt), each nucleotide with a different probability $p_i$ of being erroneous and a probability ($1-p_i$) of being correct. Our target is to obtain the probability of this sequence of having $j$ erroneous nucleotides, for $j = 0, 1, 2,\ldots, N$ (see example in **Figure 1a,b**). Statistically, our problem can be analysed as the probability distribution of the number of successes in a sequence of $N$ independent yes/no experiments with success probabilities $p_1, p_2, \ldots, p_N$. This is equivalent to the sum $S_N$ of $N$ independent Bernoulli distributed random variables $X_1, X_2, \ldots, X_N$ such that $S_N = \sum_{i=1}^{N} X_i$ , where

$$P(X_i=j) = 1-p_i \text{ for } j=0,$$

$$P(X_i=j) = p_i \quad \text{for } j=1, \qquad \text{Eq. (1)}$$

$$P(X_i=j) = 0 \quad \text{for } j>1,$$

and $P(X_i=j)$ stands for the probability of obtaining $j$ errors in nucleotide $i$. The stochastic variable $S_N$ follows a Poisson binomial distribution, from where we name the method presented here.

While the probability of obtaining a sequence with $j$ errors in a sequence, for all values of $j$, can be expressed explicitly (see Eq. (**SN1.2**) and its derivation in section **Supplementary Note 1.1**), it becomes useless in practice for moderate values of $j$. We explain here an alternative algorithm inspired by 22 that allows us to calculate the error-probability distribution $P(S_N=j)$ for all $j$ in a simple and efficient way.

First, note that if we have two random variables $Y$ and $Z$, each of them taking discrete values 0, 1, 2,..., the probability of the sum $Y+Z$ of taking value $j$ is

$$P(Y+Z=j) = \sum_{i=0}^{j} P(Y=i)P(Z=j-i). \qquad \text{Eq. (2)}$$

The algorithm results:

1. Obtain $P(X_1=j)$ from Eq. (1). Let $U = X_1$.

2. For $i = 2, 3,\ldots, N$, the distribution is obtained by following (a-c) recursively.

   (a) Calculate $P(X_i=j)$ from Eq. (1).

   (b) Calculate $P(Y+Z=j)$ from Eq. (2), being $Y = U$ and $Z = X_i$.

   (c) Let $U = Y+Z$.

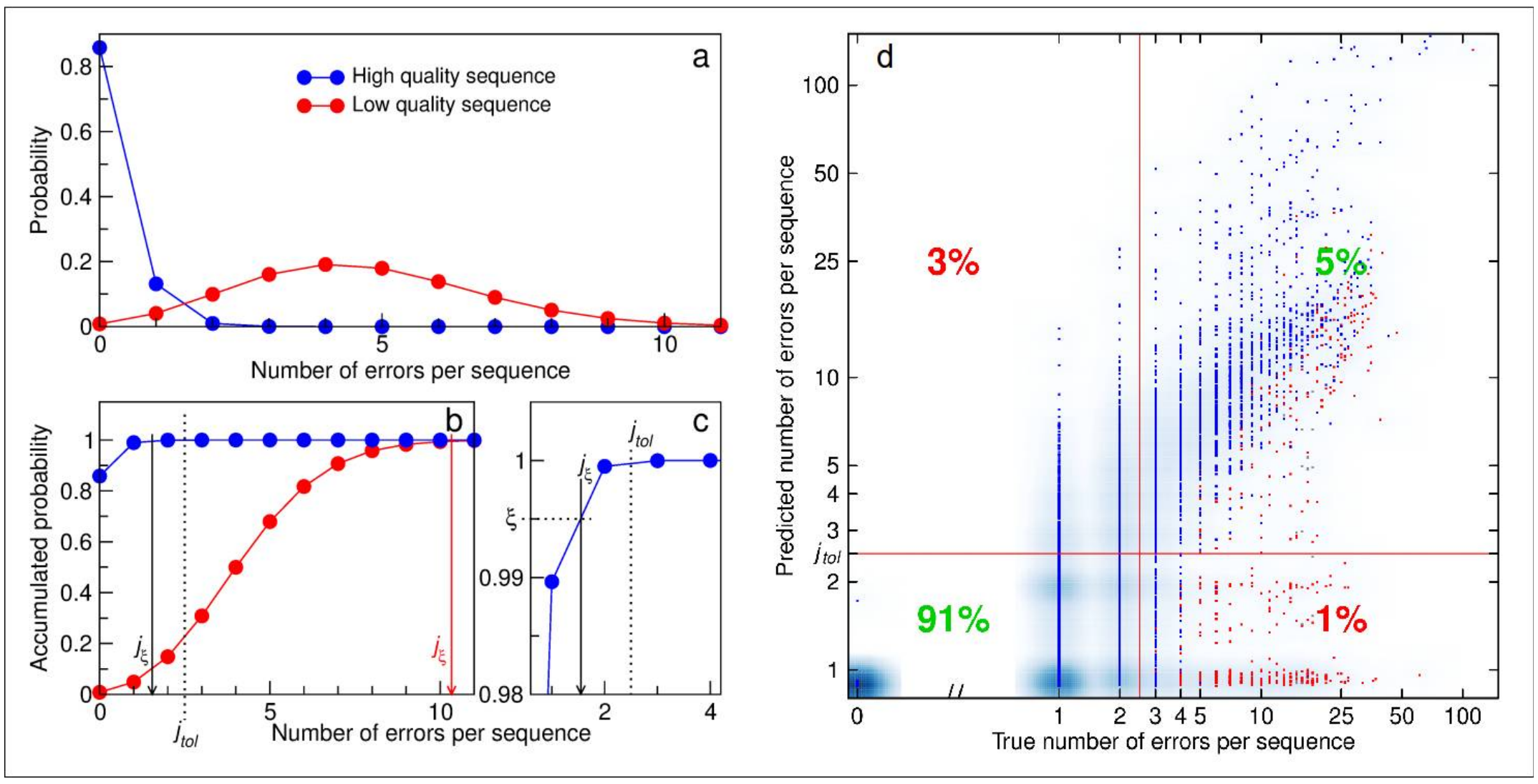

**Figure 1: Poisson binomial filtering accurately discriminates between good and erroneous sequences.**

(**a, b**): Error probability distribution (**a**) and accumulated error probability distribution (**b**) of two example nucleotide sequences, as calculated from their quality scores by the Poisson binomial filtering algorithm. $j_\xi$ stands for the 99.5th percentile of the error probability (i.e. a sequence has a probability ξ = 0.995 of having less than $j_\xi$ errors). $j_{tol}$ is the maximum tolerable number of errors (1% of the sequence length in our case). Sequences with $j_\xi > j_{tol}$ are discarded in the filtering step.

(**c**) Zoom of (**b**), sketching the calculation of $j_\xi$ for the high quality sequence.

(**d**) Comparison between the number of errors $j_\xi$ predicted by the Poisson binomial algorithm and the true number of errors for all sequences from the Even1T mock community dataset. Dots represent unique sequences. True mock community sequences are plotted in blue, contaminant sequences are plotted in gray, and chimeric sequences are plotted in red. The blue background represents sequence abundance (note that few unique sequences may have a high number of representatives, and vice versa). Red lines indicate our error cut-off of 2.5 errors per sequence ($j_{tol}$). The plot is thus divided in four quadrants corresponding to correctly retained sequences (lower left), correctly discarded sequences (upper right), incorrectly discarded sequences (upper left) and incorrectly retained sequences (lower right). The percentage of true mock community sequences present on each quadrant is also indicated. Poisson binomial filtering correctly classified 96% of the non-chimeric/non-contaminant sequences present in the Even1T dataset. The graph is plotted in logarithmic scale (the 0 in the x-axis is added for clarity).

3. The exact probability for the sequence under study of having $j$ errors, $P(S_N=j)$, is given by $U$ when $i = N$.

4. The steps (1-3) must be repeated for $j = 0, 1, 2, \ldots, j_{max}$, where $j_{max}$ is the lowest value of $j$ that satisfies $\sum_{r=0}^{j} P(S_N=r) \geq \xi$ and $0 < \xi < 1$ is a confidence coefficient (in our case $\xi = 0.995$). Let $j_\xi$ be the number such that the sequence has a probability $\xi$ of having less than $j_\xi$ errors. It is obtained interpolating the accumulated error probability of the sequence between the values $r = j_{max}-1$ and $r = j_{max}$ to obtain its exact value in $r = j_\xi$. A linear interpolation yields

$$j_\xi = j_{max} - 1 + \frac{\xi - \sum_{r=0}^{j_{max}-1} P(S_N = r)}{P(S_N = j_{max})} .$$

5. Let $j_{tol}$ be the maximum tolerable number of errors per sequence, that is, the maximum number of errors allowed for a correct clustering. In our calculations we have fixed $j_{tol}$ = 2.5. The sequence under study is discarded if $j_\xi > j_{tol}$, and accepted as correct otherwise (**Figure 1b,c**). At this moment, the calculation for this particular sequence is finished, and it is time to repeat the whole algorithm for the rest of the sequences of the population.

Finally, as our problem corresponds to the sum of *N* binomial distributions of probabilities $p_i$ and number of trials *n*=1, it can be approximated to a Poisson distribution as far as *N* is high and $p_i$ <<1. The Poisson approximated probability for the sequence under study of having *j* errors, $P(S_N=j)$, becomes

$$P(S_N = j) = \frac{\lambda^j \exp(-\lambda)}{j!} \text{ , where } \lambda = \sum_{i=1}^{N} p_i .$$

A more detailed explanation of the Poisson binomial filtering algorithm presented above and its Poisson approximation can be found in **Supplementary Note 1**.

*Algorithm implementation*

Both C and Python implementations of the Poisson binomial filtering algorithm are available in GitHub (http://github.com/fpusan/moira).

*The moira filtering pipeline*

The script *moira.py* contains an implementation of the Poisson binomial filtering algorithm and performs the following tasks:

- If required, it assembles contigs from paired reads (*--paired*). The assembler is an implementation of mothur make.contigs command (http://www.mothur.org/wiki/Make.contigs), and includes a modified version of the Needleman-Wunsch global aligner and a consensus sequence constructor. Our implementation also returns consensus quality scores, which are simply the highest quality scores for each position of the alignment.

- It collapses identical sequences and chooses the one with the highest quality as the group representative for filtering (*--collapse*). We assumed that, in spite of differences in quality, identical sequences should have the same origin, as it is unlikely that two biologically unrelated sequences become identical due to sequencing errors. Thus, if one of them has good quality, the rest should be considered as true biological sequences and be allowed into the final dataset. We have demonstrated that collapsing sequences prior to quality filtering actually helps to mitigate an important source of taxonomic bias during sequence processing (**Supplementary Note 2**).

- It truncates sequences to a fixed length (--*truncate*), discarding the sequences that are smaller than the cut-off.

- It calculates the number of errors of each remaining sequence, with a given confidence coefficient (--*alpha*) and discards the ones that have more errors per nucleotide than the specified cut-off (--*uncert*). The *alpha* confidence coefficient is defined as 1 - $\xi$, and represents the probability of underestimating the errors present on a given sequence.

The *moira.py* script can be downloaded from GitHub (http://github.com/fpusan/moira).

*16S Mock Community data*

Two synthetic mock microbial communities designed by the Human Microbiome Project (23, http://www.hmpdacc.org/HMMC) were used for evaluating the different filtering methods. Genetic DNA from 22 different organisms (20 bacterial, 1 archaeal and 1 eukaryotic) was mixed in known amounts, based on qPCR of the small subunit (SSU) rRNA gene, in order to generate two different mixtures: an Even mock community, in which there is a similar amount of SSU rRNA copies for each organism, and a Staggered mock community, in which the amounts of SSU rRNA of each organism are different.

The data used in this study come from publicly available libraries generated by sequencing the Even and Staggered mock communities with the Roche 454 GS FLX Titanium, the Illumina MiSeq and the IonTorrent PGM platforms. References for all the datasets used in this study are given in **Supplementary Note 3.**

*Validation of Poisson binomial filtering on Mock Community data*

The script *moira.py* was used to predict the number of errors present on each sequence for all the six Roche 454 GS FLX Titanium, the four Illumina MiSeq and the two IonTorrent PGM mock community datasets. For the MiSeq datasets, contigs were first assembled from paired-end reads by applying the --*paired* flag. The --*alpha* parameter, which indicates the probability of a read having more errors than reported, was left as its default value of 0.005. Identical reads were collapsed and the sequence with the smallest number of errors was chosen as the group representative, as described above. These predicted values were compared to the true number of errors of each sequence, which was obtained by using the mothur command *seq.error*. Briefly, the sequences were aligned to a reference database made up from the true biological sequences present in the mock community (which can be found in http://www.mothur.org/wiki/454_SOP). Sequences with less than 80% alignment coverage were discarded at this step. The resulting alignment was then used to determine the true number of errors present on each sequence, as well as whether that sequence was chimeric or not. Non-chimeric sequences that nevertheless showed less than 95% similarity to their best hit in the mock reference database were aligned again against mothur's SILVA 16S reference alignment (version 98). In case said sequence showed a pairwise identity and an alignment coverage equal or greater to 95% to any sequence in the 16S reference alignment, it was considered to be a contaminant.

*Quality filtering of 16S reads*

- Usearch: Trimming of reads by quality values was performed by using the USEARCH *fastq_filter* command, as employed by 15. Reads (for 454/IonTorrent data) or contigs (for paired Illumina data) were truncated at the first position with a quality score below 15 (--*fastq_trunqual* 15). After that, sequences were truncated to a length of 250 nucleotides (200 nucleotides for IonTorrent data), and sequences smaller than 250 nt (200 nt for IonTorrent data) were discarded (--*fastq_trunclen* 250/200).

We also tested a different method implemented in the USEARCH *fastq_filter* command, as suggested in the author's web page (http://drive5.com/usearch/manual/uparse_cmds.html). Briefly, reads (for 454 data and IonTorrent) or contigs (for paired MiSeq data) with more than 0.5 expected errors (--*fastq_maxee* 0.5) were discarded. After that, sequences were truncated to a length of 250 nt (200 nt for IonTorrent data), and sequences smaller than 250 nt (200nt for IonTorrent data) were discarded (--*fastq_trunclen* 250/200).

- mothur: Denoising of 454 and IonTorrent reads was performed using the mothur command *shhh.flows*, which is an implementation of the PyroNoise algorithm, as recommended in the mothur SOP (http://www.mothur.org/wiki/454_SOP; http://www.mothur.org/wiki/Ion_Torrent_sequence_analysis_using_Mothur). After denoising, mothur command *trim.seqs* was used to truncate the denoised sequences to a length of 250 nt, and to discard sequences smaller than 250 nt (200nt in both cases for IonTorrent data).

Additionally, paired Illumina reads were assembled and filtered according to mothur's MiSeq SOP (http://www.mothur.org/wiki/MiSeq_SOP). After filtering, mothur command *trim.seqs* was used to truncate the contigs to a length of 250 nt, and to discard sequences smaller than 250 nt.

- QIIME: QIIME's script *split_libraries_fastq.py* was used to filter Illumina forward reads, as recommended by the authors (-*r* 3 -*p* 0.75 -*q* 3 -*n* 0) in 21.

- Poisson binomial filtering: The script *moira.py* was used to perform Poisson binomial filtering on 454/IonTorrent reads or contigs assembled from Illumina paired reads (--*paired*), as described above. Before filtering, sequences or contigs were truncated to 250 nt, and the sequences smaller than 250 nt (200nt in both cases for IonTorrent reads) were discarded (--*truncate* 250/200). Identical 454/IonTorrent reads or Illumina contigs were clustered together prior to quality control (--*collapse*) and the sequence with the highest quality was chosen as the group representative for quality control. 0.01 or less errors per nucleotide were tolerated (--*uncert* 0.01) with a 0.005 chance of error underestimation (--*alpha* 0.005).

For each method, paired Illumina reads were assembled as recommended by its authors.

Note that, for consistency, we have chosen the 250 nt cut-off recommended by 15 as the fixed length for the rest of the filtering methods (except for QIIME, which works with unpaired Illumina reads that had a fixed length of 250 nt on their own), for the 454 and Illumina datasets. Since read length may have an effect in clustering and OTU accuracy, we believe that equalizing it results in more valid comparisons between the different filtering methods. In a similar fashion, the 200 nt cutoff proposed in http://www.brmicrobiome.org/#!16sprofilingpipeline/cuhd was applied to all the filtering methods for the IonTorrent datasets.

The full list of commands used for each method can be found in **Supplementary Note 6**.

*Common processing pipeline for the filtered reads*

Regardless of the filtering method, the filtered sequences were subjected to a common pipeline based in mothur's recommended SOP that included the following steps:

- Sequence alignment to mothur's Silva Reference Alignment.

- Optimization of the alignment space by removing the sequences that failed to align correctly.

- Pre-clustering of similar sequences.

- Removal of chimeras with UCHIME.

- Taxonomic classification and removal of non-bacterial and unclassified sequences.

- Library size standardization (see below).

- Clustering of the remaining sequences using mothur's default average neighbour algorithm, with an OTU distance cut-off of 0.03.

- Accuracy classification of the resulting OTUs (see below).

For each sample, the libraries obtained after filtering the raw reads with the different methods were standardized to a similar size by random sub-sampling. Total number of retrieved OTUs and singletons, as well as accuracy assessment of the OTU representative sequences, were obtained by averaging the results from 100 independent rounds of random library size standardizations followed by clustering of the resulting reads.

The full list of commands can be found in **Supplementary Note 6**.

*OTU accuracy assessment on mock communities*

The accuracy of the obtained OTU representative sequences was evaluated by aligning them to a reference database made up from the true biological sequences present in the sample, as previously described by 15. Sequence alignment was performed with mothur *align.seqs* command. If the pairwise identity of an OTU representative sequence to any sequence in the reference database was 100%, the OTU was classified as "Perfect". If the pairwise identity was smaller than 100%, but greater or equal to 99%, the OTU was classified as "Good". If the pairwise identity was smaller than 99% but greater or equal to 97%, the OTU was classified as "Noisy". If the pairwise identity was lower than 95% the OTU representative sequence was aligned to mothur's SILVA bacterial 16S reference alignment (version 98). If said sequence showed a pairwise identity and an alignment coverage equal or greater to 95% to any sequence in the 16S reference alignment, the OTU was classified as "Contaminant". When none of the above conditions applied, the OTU was considered to be the result of either an undetected chimera or a mock community sequence with more than 3% errors, and was classified as "Other".

*OTU accuracy assessment on environmental communities*

OTU representative sequences from environmental communities were aligned with mothur's SILVA bacterial 16S reference alignment. For each dataset and filtering method, the average similitude of the OTU representative sequences to their best hits in the SILVA alignment was calculated. This was taken as an indicator of the overall accuracy of the resulting OTUs, under the assumption that sequencing errors are more likely to decrease OTU similitude to known sequences than to increase it.

References for all the environmental datasets used in this study are given in **Supplementary Note 3.**

*Assessment of the taxonomic bias caused by the different filtering methods*

Taxonomic bias was assessed by comparing the taxonomic composition of the sample before and after performing quality filtering. Sequences were classified by using the *classify.seqs* command implemented in mothur and mothur's RDP 16S rRNA reference database (version 9). Then, taxonomic composition was obtained by calculating the proportion of sequences that were assigned to each phylotype at the genus level with an 80% confidence cut-off (40% for the environmental communities). Finally, taxonomic bias was calculated as the Bray-Curtis dissimilarity between the filtered and unfiltered sequence communities. In the 454 and IonTorrent libraries from the environmental communities, a high proportion of sequences did not get classified at the genus level. Therefore, the taxonomic composition of those libraries was instead calculated at the class level.

## Results

We validated the Poisson binomial filtering algorithm and compared it with the different filtering approaches recommended by the authors of mothur (8, 13, 20), USEARCH-UPARSE (10, 15) and QIIME (21) by quality-filtering datasets obtained by sequencing different mock and environmental microbial communities with the Roche 454 GS FLX Titanium, the Illumina MiSeq and the IonTorrent PGM platforms. In order to evaluate the different methods on equal grounds, filtered reads were processed with a common downstream pipeline that included chimera-filtering with UCHIME (24), sample size standardization and OTU clustering.

### *Poisson binomial filtering accurately discriminates between good and erroneous sequences*

When applying our default cut-off of 1% errors allowed per sequence, our algorithm accurately classified 96% of the mock community sequences from the Even1M dataset (**Figure 1d**). 3% of the sequences were incorrectly discarded while, remarkably, only 1% of the sequences were incorrectly retained. Moreover, most of those incorrectly retained sequences had only 3 true errors (1.2% errors per sequence), meaning that they would likely cluster correctly when applying the standard 3% OTU distance cut-off. The rest of the Illumina datasets rendered similar results. The accuracy of our method was slightly lower for the 454 and IonTorrent datasets, but it nevertheless resulted in a minimum of 88% (for 454) and 79% (for IonTorrent) correctly classified sequences. (**Supplementary Figure SN4.1**).

### *Performance of the different filtering methods on mock community datasets*

Publicly available datasets from even and staggered mock communities from the Human Microbiome Project (23) were filtered with PBF, mothur, USEARCH and QIIME (**Figure 2, Supplementary Note 4**). These artificial communities contain known amounts of 16S rRNA gene copies from 20 different bacterial organisms. The fact that both the qualitative and quantitative composition of the samples are known beforehand allowed us to thoroughly compare the effects of the different filtering methods in terms of OTU accuracy, alpha diversity and community composition. OTU accuracy was defined as the maximum similarity of its representative sequence to the 16S sequences of the microorganisms used to build the mock community, as previously described in 15. We were also interested in determining how the different filtering processes affected the observed community composition. The taxonomic bias in community composition caused by any given filtering method was calculated as the Bray-Curtis dissimilarity between the raw and the filtered datasets, after taxonomically classifying their reads down to the genus level.

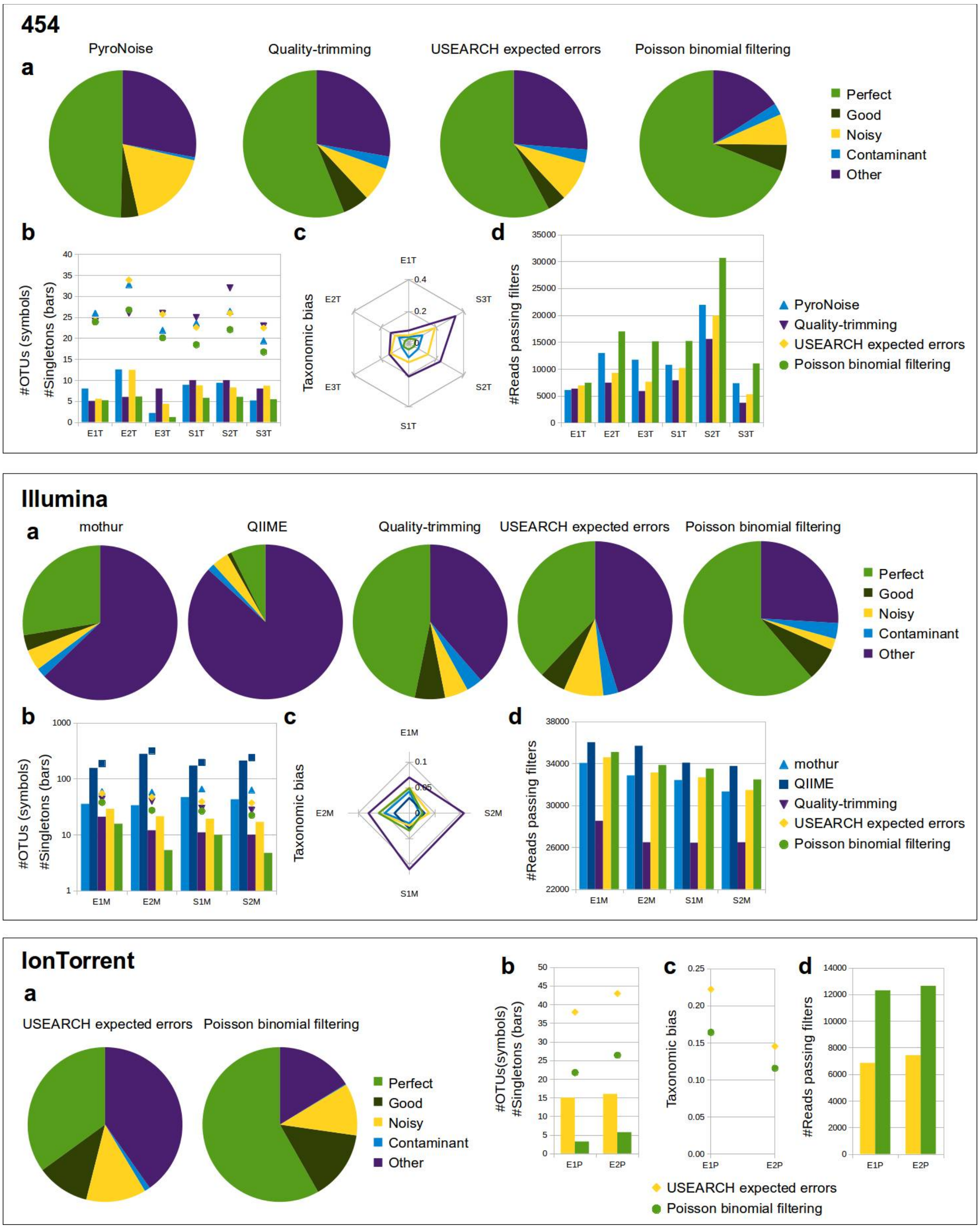

**Figure 2: Comparison of filtering methods on 16S mock communities sequenced with the 454 GS FLX Titanium, Illumina MiSeq platforms and IonTorrent PGM platforms.**

**(a)** Pie charts constructed by averaging the fraction of OTUs on each accuracy category along the six 454 or the four Illumina samples.

**(b , d)** Number of singletons (**b,** bars), total species (**b**, symbols) and reads **(d)** retrieved after filtering the raw reads with the different methods and performing chimera removal and clustering with a common pipeline. OTU and singleton numbers were obtained by averaging the results from 100 independent library size standardizations.

**(c)** Taxonomic bias caused by the different filtering methods, measured as the Bray-Curtis dissimilarity between the raw and the filtered read communities.

In the even datasets, which contain the same number of 16S rRNA gene copies for each organism, all methods resulted in more than 20 OTUs after clustering. This was not surprising, since contaminations, PCR errors and sequencing errors were expected to inflate the observed diversity. In the staggered communities, in which the number of 16S rRNA gene copies varied by several orders of magnitude between the different organisms, the observed diversity was generally lower, due to some species being present at very low abundances. The total number of reported OTUs greatly varied between filtering methods, with Poisson binomial filtering consistently resulting in values that were the closest to the true diversity of the samples.

PBF also produced the highest proportion of accurate OTUs in all the 16S mock datasets for both sequencing platforms, while minimizing the number of singletons and spurious OTUs retrieved (**Figure 2a,b**). In the 454 and IonTorrent datasets it also discarded the smallest number of reads and resulted in the smallest taxonomic bias (**Figure 2c,d**). In the Illumina datasets QIIME retrieved a larger number of reads, while both QIIME and mothur caused smaller taxonomic biases than our method. (**Figure 2c,d - Illumina**). However, we believe that this was the result of a too shallow filtering by mothur and QIIME, since both methods produced a remarkably lower proportion of accurate OTUs and a larger number of OTUs and singletons (**Figure 2a,b - Illumina**). QIIME produced an especially high number of spurious OTUs, a fact that has also been discussed elsewhere (15). Their pipeline (21) deals with this problem by applying a post-hoc OTU size cut-off at the cost of sensitivity. Nonetheless, our results show that, even after the removal of singletons from the QIIME-filtered dataset, their number of OTUs would exceed that of the dataset filtered with our method, including singletons (**Supplementary Table SN4.4**).

The two filtering algorithms included in the USEARCH suite showed an intermediate performance in terms of the number and accuracy of the OTUs retrieved for both the 454 and Illumina platforms. Quality trimming yielded the smallest number of reads and resulted in the highest taxonomic bias, which supports the idea that over-stringent filtering may lead to undesirable effects. In the IonTorrent datasets, USEARCH filtering performed below Poisson binomial filtering for all the studied bechmarks (**Figure 2 - IonTorrent**). Finally, the mothur implementation of the PyroNoise algorithm (12) showed lower OTU accuracy than the other methods tested for filtering 454 reads. It has been previously described that the denoising process can introduce minor alterations in the original reads (25), a phenomenon that might explain these results. It must be noted that, albeit a pipeline for filtering IonTorrent reads with PyroNoise has been described, the IonTorrent mock community datasets were only available in Fastq format (Stephen Salipante, personnal communication), which precluded the use of flowgram denoising algorithms. However, this limitation was not present for the environmental datasets, and a comparison of quality filtering algorithms for IonTorrent datasets that includes PyroNoise can therefore be found in **Supplementary Figure SN5.3**.

*Performance of the different filtering methods on environmental datasets*

The performance of the different filtering methods was also evaluated by quality-filtering publicly available datasets obtained by sequencing environmental communities (**Supplementary Note 5**). The results were similar to the ones obtained with the mock communities, with Poisson binomial filtering being the most consistent method in producing the smallest number of OTUs and singletons. Additionally, the OTUs obtained with PBF were overall the most similar to the 16S sequences present in the SILVA 16S reference alignment (26), which suggests that they contained the smallest number of errors. In the environmental 454 datasets, PyroNoise showed better results than in the 454 mock communities, but did it in an irregular fashion, especially in terms of OTU accuracy (**Supplementary Figure SN5.1d**). This inconsistency may be again due to the alteration of the original reads, and suggests that PyroNoise requires a finer parameter optimization than other approaches in order to be fully effective. In the environmental IonTorrent datasets PyroNoise discarded the smallest number of reads, but resulted in the highest number of singletons and OTUs, which also beared the least similarity to the reference alignment. USEARCH showed an intermediate performance between PyroNoise and Poisson binomial filtering (**Supplementary Figure SN5.3**). Finally, in the environmental Illumina datasets all filtering methods showed a similar behaviour to that in the mock communities (**Supplementary Figure SN5.2**).

*Quality-filtering is an additional source of taxonomic bias in microbial ecology studies*

Even though the major sources of taxonomic biases in marker-gene-based studies are often related to PCR and library construction (27, 28, 29), the read filtering process can greatly exacerbate this problem (**Supplementary Note 2**). We found significant biases in length and quality distribution between raw reads coming from different taxa in the mock 454 datasets (**Figs. 3 a,b,c**). Trimming them to a fixed length generated an artificial enrichment of the taxa with longer reads (**Figure 3b**), but since there is a decrease in quality at the end of 454 reads (see 15), it also resulted in a lower average read quality for the taxa with smaller raw reads (**Figure 3d**). This led to the generation of further taxonomic bias during the quality-filtering step (**Figure 3f**, **Supplementary Note 2**). Similar biases have been previously found in IonTorrent reads (30), and were confirmed during this study (**Supplementary Note 2)**. Biases in read quality distribution between different taxa were also found for the mock Illumina datasets, although to a lesser extent. We solved this problem by collapsing identical reads and choosing the one with the highest quality as a representative for filtering, in order to decide whether the whole group was discarded or allowed into the filtered dataset. This procedure reduced the effect of quality distribution biases, as even low abundance sequences are expected to have a high quality representative. Our solution rendered similar quality distributions for the different taxa, even after length trimming (**Figure 3e,f**), and significantly lower taxonomic biases than other filtering approaches, especially for 454 data (**Figure 2c**). Every method that relies on quality scores for sequence filtering will be affected by this source of bias. We therefore propose the approach described above as a general solution to this problem, since its simplicity makes it very easy to integrate into any filtering pipeline.

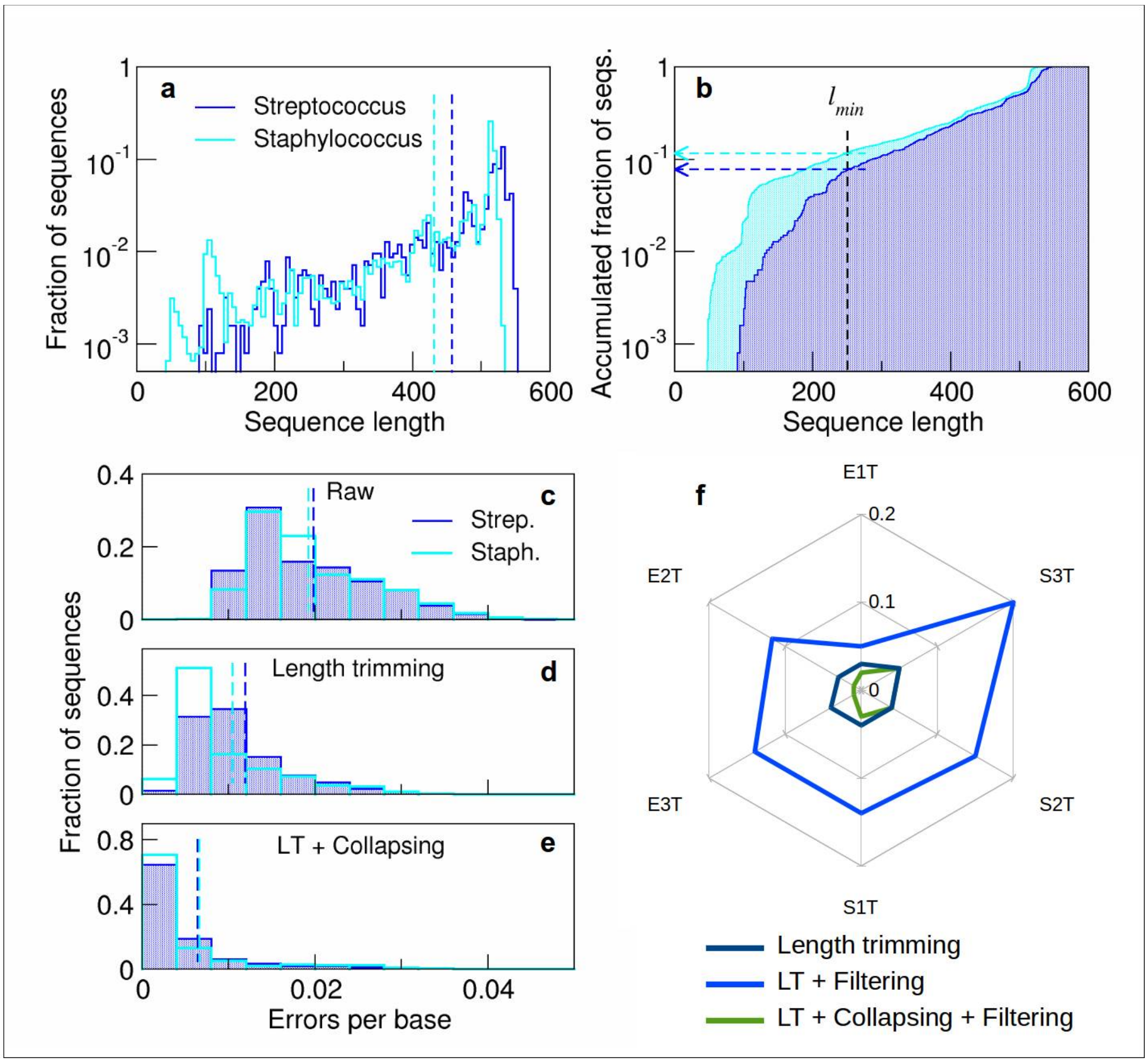

**Figure 3: Addressing the taxonomic bias generated during the pre-processing and quality filtering of raw sequences.**
**(a, b)**: Raw reads from Streptococcus and Staphylococcus, the two most abundant genera in sample Even3T, show different length distributions. The dashed vertical lines in (a) indicate the average read lengths. The arrows in (**b**) indicate the fraction of reads from each taxon removed after discarding sequences shorter than lmin= 250 nt.
**(c, d, e)**: Errors per base distributions of Streptococcus and Staphylococcus reads in the (**c**) raw dataset, (**d**) after trimming the reads to 250 nt and discarding the ones shorter than the cut-off, and (**e**) after collapsing the trimmed reads. The dashed lines indicate average errors per base. Note that length trimming substantially increases the difference between the Streptococcus and Staphylococcus error distributions (**d**) when compared to that of the raw reads (**c**). Filtering at this point would cause a 56.2% overrepresentation of Streptococcus versus Staphylococcus (see text and **Supplementary Note 2**). Collapsing identical reads prior to filtering solves this problem **(e)**, reducing the overrepresentation to 1%.
**(f)**: Compositional bias generated during the pre-processing and filtering of the six 454 mock community samples, measured as the Bray-Curtis dissimilarity between the raw and the processed read communities. This shows that results in (**c**, **d**, **e**) can be generalized to all the taxa present in all the samples.

## Discussion

In this work we have presented and validated the Poisson binomial algorithm for filtering sequence reads based on their error probability distributions. We have also demonstrated that Poisson binomial filtering is especially useful in the context of gene-marker-based studies, such as the study of microbial populations by amplifying and sequencing their 16S rRNA gene.

We compared our algorithm with other five quality-filtering methods that are included as defaults in mainstream pipelines such as mothur, QIIME or USEARCH, by analysing mock and environmental datasets generated with three different sequencing platforms. Our results show that, when coupled to a standard analysis pipeline that included chimera removal and clustering, PBF proved to be the most accurate algorithm for filtering marker-gene sequences. While retaining a large number of sequences, it also resulted in OTUs that were the closest to the true biological species present in the studied samples, and minimized the generation of spurious diversity and taxonomic biases.

Remarkably, this algorithm does not rely on any particular error model. Instead, it just derives the error probability distribution of a given sequence from the quality scores of its individual bases. To our knowledge, it is the first non-heuristic method to do so. The only assumption that our algorithm makes (which is shared with any other approach that utilizes quality scores) is that, for any given sequencing platform, the quality scores obtained during base calling will truly represent the probabilities of that base being wrong. This conceptual simplicity is one of its main advantages: as long as accurate quality scores are provided, Poisson binomial filtering will work in any present or future sequencing platform, with no need for further modifications.

In practice, quality-score calling ultimately depends on the sequencing platform manufacturer, and its accuracy is also influenced by the choice of primers and library preparation methods (31, 32). Nonetheless, we have shown that, for the three sequencing platforms studied in this work, Poisson binomial filtering was able to correctly discriminate between good and erroneous sequences based solely on quality score information.

The fact that our method only relies on quality scores means that it will only account for sequencing errors, but not other errors such as PCR substitutions. However, it has been described that sequencing errors are responsible for the majority of singletons generated in molecular ecology studies (15, 17). PCR chimeras are other source of spurious diversity, but dedicated algorithms such as UCHIME are able to accurately detect them.

During the course of this research, we have also focused on a source of taxonomic bias that may have affected the results of many molecular ecology studies. Most of the methods used for filtering and analysing marker-gene reads operate under the implicit (or even explicit, see 17) assumption that the probability of having $k$ errors is the same for all sequences, regardless of their origin. However, sequences from different taxa may have different length (for 454 and IonTorrent) and quality (for 454, IonTorrent and Illumina) distributions. This leads to the artificial enrichment of some taxa versus others during the quality filtering step, potentially compromising the quantitative interpretation of molecular ecology results obtained by high-throughput sequencing of marker-gene sequences. These biases are likely originated during base/quality calling: for instance, 454 reads show a systematic decrease in quality after homopolymer regions (33), which will penalize the taxa with longer homopolymer stretches on its 16S gene. We have nonetheless demonstrated that collapsing identical reads before the quality-filtering step greatly mitigates this issue.

In summary, the methodologies presented in this work substantially improve the existing filtering approaches in terms of OTU accuracy, observed alpha diversity and observed community composition, delivering a more faithful representation of the original microbial communities present in the studied samples. Our algorithm is fast, easy to implement, and works for every sequencing platform constituting a valuable addition to all the existing pipelines for analysing microbial ecology data.

**Supplementary information** is available at https://github.com/fpusan/doc**.**

## Funding

This work was supported by the European Research Council [250350]; Subdirección General de Proyectos de Investigación of Spanish Ministerio de Economía y Competitividad [YA2011-24803, FIS2011-27569, FIS2014-57686-P] and the Spanish Comunidad de Madrid [MODELICO-CM-S2009ESP-1691]. Computational resources were provided by the Data Intensive Academic Grid, which is supported by the United States National Science Funcation [0959894]. F.P.S. was supported by a JAE-pre fellowship from the Spanish Consejo Superior de Investigaciones Científicas (CSIC).

## Acknowledgements

The authors thank G. Ackermann, N.A. Bokulich, J.G. Caporaso and J.R. Rideout for facilitating access to the MiSeq mock community datasets, S. Salipante for providing information on the IonTorrent mock community datasets, P.S. Schloss for his insights into mothur's implementation of the Needleman-Wunsch algorithm, and A. Arce-Rodríguez, M.J. Gómez, J. Iranzo, S. Lincoln, F. López de Saro and S. Manrubia for useful feedback and discussions of the draft manuscript.

## Author contributions

F.P.S. and J.A. conceived the algorithm. J.A. developed the analytical work. F.P.S. and V.P. conceived the experiments. F.P.S. implemented the algorithm and designed and performed the experiments. F.P.S., J.A. and V.P. participated in the motivation and discussion of the results and contributed to the writing of the manuscript.

## Competing financial interests

The authors declare no competing financial interests.